\documentclass[preprint,10pt]{elsarticle}
\usepackage{graphicx}
\usepackage{amssymb}
\usepackage{amsmath}

\usepackage{color}

\allowdisplaybreaks

\newcommand{\gt}{\texttt}

\newcounter{bla}

\journal{Computer Physics Communications}

\begin{document}

\begin{frontmatter}

\title{CIJET: a program for computation of jet cross
sections induced by quark contact interactions at hadron colliders}

\author{Jun Gao\corref{}}

\cortext []{\textit{E-mail address:} jung@smu.edu}

\address{Department of Physics, Southern Methodist University,
Dallas, TX 75275-0175, USA}

%--------1---------2---------3---------4---------5---------6---------7---------8
\begin{abstract}
We describe CIJET1.0, a Fortran program that aiming for the calculation
of single-inclusive jet or dijet production cross sections induced by quark
contact interactions from new physics at hadron colliders, up to next-to-leading
order in QCD. It covers various contact
interactions with different chiral and color structures. The code is designed in a way
that could be used for fast calculations with arbitrary parton distribution functions
based on interpolations of the QCD coupling constant and parton distribution functions.
\end{abstract}


\end{frontmatter}

%%
%% Start line numbering here if you want
%%
% \linenumbers

% Computer program descriptions should contain the following
% PROGRAM SUMMARY.

\noindent
{\bf PROGRAM SUMMARY}
  %Delete as appropriate.

\begin{small}

\noindent {\em Manuscript Title:} 
CIJET1.0: a program for computation of jet cross
sections induced by quark contact interactions at hadron colliders
 \\
{\em Authors:} Jun Gao\\
{\em Program Title:} CIJET 1.0\\
{\em Journal Reference:}                                      \\
  %Leave blank, supplied by Elsevier.
{\em Catalogue identifier:}                                   \\
  %Leave blank, supplied by Elsevier.
{\em Preprint number:} SMU-HEP-13-08 \\ 
{\em Licensing provisions:} none\\
  %enter "none" if CPC non-profit use license is sufficient.
{\em Programming language:} Fortran;
with some included libraries coded in C\\
{\em Computer:} all\\
  %Computer(s) for which program has been designed.
{\em Operating system:} any UNIX-like system\\
  %Operating system(s) for which program has been designed.
{\em RAM:} $\sim$ 300 MB\\
  %RAM in bytes required to execute program with typical data.
% {\em Number of processors used:} \\
  %If more than one processor.
{\em Supplementary material:}                             \\
  % Fill in if necessary, otherwise leave out.
  % Please give some freely chosen keywords that we can use in a
  % cumulative keyword index.
{\em Classification:} 11.1\\
  %Classify using CPC Program Library Subject Index, see (
  % http://cpc.cs.qub.ac.uk/subjectIndex/SUBJECT_index.html)
{\em Nature of problem:} Calculation of single-inclusive jet and
dijet cross sections induced by quark contact interactions up to next-to-leading
order in QCD at hadron colliders.
\\
{\em Solution method:} Using the VEGAS Monte Carlo sampling and
subtracted matrix element to generate weight grids and store them
in a table file that allows for later interpolations on arbitrary
parton distribution functions.
\\
{\em Running time:} Depends on details of the calculation and the
required numerical accuracy. \\
%--------1---------2---------3---------4---------5---------6---------7---------8

\end{small}

%% main text
\section{Introduction}
Jet production at hadron colliders provides an excellent opportunity
to test perturbative QCD (PQCD) and to search for possible new
physics (NP) beyond the Standard Model (SM) over a wide range of
energy scales. Invariant mass distributions of the
dijets~\cite{Chatrchyan:2011ns}, dijet angular
distributions~\cite{Khachatryan:2011as,Aad:2011aj,Chatrchyan:2012bf},
inclusive jet $p_T$ distribution~\cite{:cmspt}, and other jet observables
at the LHC~\cite{:2010wv} have already
extended current searches for quark compositeness, excited quarks,
and other new particle resonances toward the highest energies
attainable. Among all these measurements, the inclusive jet $p_T$ distribution and
dijet angular distribution show a great sensitivity to possible quark contact
interactions (CIs) induced by new physics models. In the SM, Quantum
Chromodynamics (QCD) predicts the jets are preferably produced in large
rapidity region, via small angle scattering in $t$-channel processes. Also the
production rates fall off rapidly in large $p_T$ or large dijet invariant mass region.
On the contrary, the jet production induced by quark contact interactions is
expected to be much more isotropic and fall off slower as the increasing of
jet $p_T$ or dijet invariant mass, and thus the distributions at the LHC could
be largely modified.

The measurement of quark contact interactions has been used to set
limits on the quark composite models which have been studied
extensively in the literature~\cite{Eichten:1983hw,Lane:1996gr}. It
is assumed that quarks are composed of more fundamental particles
with new strong interactions at a compositeness scale $\Lambda$,
much greater than the quark mass scales. At the energy well below
$\Lambda$, quark contact interactions are induced by the underlying
strong dynamics, and yield observable signals at hadron colliders.
The newest bounds of $\Lambda$ at the $95\%$ confidence level (C.L.)
from the CMS collaboration are around $10\,{\rm
TeV}$ based on $2.2\,{\rm fb}^{-1}$
collected data of the dijet angular measurement~\cite{Chatrchyan:2012bf} and $5\,{\rm fb}^{-1}$
data of the single-inclusive jet measurement~\cite{:cmspt}. Previous limits from the Tevatron and LHC can be
found in Refs.~\cite{Khachatryan:2011as,Aad:2011aj,Abe:1996mj,Collaboration:2010eza}.

In our earlier studies~\cite{Gao:2011ha}, we presented the theoretical calculations of the
next-to-leading (NLO) order QCD corrections to the dijet production induced by
quark CIs, which are used by the experimentalists to set constraints on the quark
compositeness scales~\cite{Chatrchyan:2012bf}. In this work we summarize the numerical program used
for the dijet calculation with extensions for the calculation of single-inclusive
jet production, which will be used in the upcoming measurement~\cite{:cmspt}. One common issue
of the experimental analyses at the hadron colliders is that, they require massive
repetitive calculations using different parton distribution functions (PDFs) to estimate the theoretical
uncertainties. And fast calculation algorithms for arbitrary PDFs, like the implementations 
of FastNLO~\cite{Kluge:2006xs} and APPLGRID~\cite{Carli:2010rw} are highly desirable.
Thus in our program we further provide a
fast interpolation modular that allows for the calculations of using arbitrary
PDFs within seconds while maintaining the desirable numerical accuracy.    

This document is structured as follows: Section \ref{s1} briefly summarizes
the theoretical model and calculations. Section \ref{s2} describes the fast
interpolation algorithm used. And Section \ref{s3}-\ref{s5} show the inputs, running,
and outputs of the program. Section \ref{s6} shows some sample results from the
program and Section \ref{s7} is a conclusion.

\section{Jet cross sections and quark contact interactions}\label{s1}
We consider a subset of quark contact interactions that are the
products of electroweak isoscalar quark currents which are assumed
to be flavor-symmetric to avoid large flavor-changing
neutral-current interactions~\cite{Lane:1996gr}. The effective
Lagrangian can be written as
\begin{equation}\label{opt} 
\mathcal{L}_{NP}=\frac{1}{2\Lambda^{2}}\sum_{i=1}^6c_{i}O_{i},
\end{equation}
where $\Lambda$ is the new physics scale, $c_i$ are Wilson
coefficients. And the operators $O_i$ in chiral basis are given by
\allowdisplaybreaks{\begin{eqnarray} O_{1} & = &
\delta_{ij}\delta_{kl}\left(\sum_{c=1}^{3}\bar{q}_{Lci}\gamma_{\mu}
q_{Lcj}\sum_{d=1}^{3}\bar{q}_{Ldk}\gamma^{\mu}q_{Ldl}\right),\nonumber \\
O_{2} & = & {\rm T}_{ij}^{a}{\rm T}_{kl}^{a}\left(\sum_{c=1}^{3}\bar{q}_{Lci}
\gamma_{\mu}q_{Lcj}\sum_{d=1}^{3}\bar{q}_{Ldk}\gamma^{\mu}q_{Ldl}\right),\nonumber \\
O_{3} & = & \delta_{ij}\delta_{kl}\left(\sum_{c=1}^{3}\bar{q}_{Lci}\gamma_{\mu}
q_{Lcj}\sum_{d=1}^{3}\bar{q}_{Rdk}\gamma^{\mu}q_{Rdl}\right),\nonumber \\
O_{4} & = & {\rm T}_{ij}^{a}{\rm T}_{kl}^{a}\left(\sum_{c=1}^{3}\bar{q}_{Lci}
\gamma_{\mu}q_{Lcj}\sum_{d=1}^{3}\bar{q}_{Rdk}\gamma^{\mu}q_{Rdl}\right),\nonumber\\
O_{5} & = & \delta_{ij}\delta_{kl}\left(\sum_{c=1}^{3}\bar{q}_{Rci}\gamma_{\mu}
q_{Rcj}\sum_{d=1}^{3}\bar{q}_{Rdk}\gamma^{\mu}q_{Rdl}\right),\nonumber \\
O_{6} & = & {\rm T}_{ij}^{a}{\rm
T}_{kl}^{a}\left(\sum_{c=1}^{3}\bar{q}_{Rci}
\gamma_{\mu}q_{Rcj}\sum_{d=1}^{3}\bar{q}_{Rdk}\gamma^{\mu}q_{Rdl}\right),
\end{eqnarray}}in which $c$, $d$ are generation indices and $i$, $j$, $k$, $l$, $a$
are color indices, and ${\rm T}^{a}$ are the Gell-Mann matrices with
the normalization ${\rm Tr}({\rm T}^{a}{\rm T}^b)=\delta^{ab}/2$.
Beside of the quark compositeness, the above interactions can also
arise from various kinds of new physics models, induced by the
exchange of new heavy resonances, such as $Z'$
models~\cite{Langacker:2008yv} and extra dimensions
models~\cite{Randall:1999ee}.

The jet cross sections (up to NLO) in each kinematic
bin, depending on the scale $\Lambda$ and Wilson 
coefficients $c_i$ at input scale $\Lambda$, can be expressed as
\begin{eqnarray}\label{fit}
\sigma_{bin}&=&\sum_{i=1}^6(\lambda_i(b_{i}+a_ir))/\Lambda^2
+\sum_{i=1}^6(\lambda_i^2(b_{ii}+a_{ii}r))/\Lambda^4 \nonumber\\
&&+\sum_{i=1,3,5}(\lambda_i\lambda_{i+1}(b_{ii+1}+a_{ii+1}r))/\Lambda^4\nonumber\\
&&+\sum_{i=1,2,5,6}(\lambda_i\lambda_{4}(b_{i4}+a_{i4}r))/\Lambda^4.
\end{eqnarray}
with $\lambda_i$ defined by $c_{i}=4\pi \lambda_{i}$, $r=\ln(\Lambda/\mu_0)$, and $\mu_0$ is an arbitrary
reference scale chosen according to the kinematic range of the bin (not the same as QCD scales).
All $a$'s vanish at leading order (LO). Since QCD interaction preserves parity, we
should have $b(a)_{1,2}=b(a)_{5,6}$,
$b(a)_{11,22,12}=b(a)_{55,66,56}$, and $b(a)_{14,24}=b(a)_{54,64}$. More details
of the theoretical calculations can be found in~\cite{Gao:2011ha}.

\section{Grid interpolation}\label{s2}
In general, dependence of the jet cross sections on the PDF and the QCD coupling constant $\alpha_s$ can be
expressed as
\begin{equation}
\sigma_{bin}=\sum_{i, j}\int dx_1dx_2f_{i/h_1}(x_1,\mu_f)f_{j/h_2}(x_2,\mu_f)\sum_p \alpha_s^p(\mu_r)
\mathcal{B}_p(x_1, x_2; \mu_f, \mu_r),
\end{equation}
where $i, j$ are the parton flavors, $p$ indicates the order of $\alpha_s$, and $\mu_{f(r)}$
are the factorization and renormalization scales. We choose a 3-dimensional grid of
$x_1-x_2-\mu_f$ with $N_x\times N_x\times N_q$ grid points, which are uniformly spaced in
$x_1^{\delta}$, $x_2^{\delta}$ and $\ln\ln{\mu_f/Q_0}$ with $\delta=0.3$ and $Q_0=0.3$ GeV.
We choose $N_x=30$ and $N_q=15$ which are found to be good enough with an interpolation
accuracy of about 0.1\%.  
Both ends of each dimension are determined according to the kinematic range of the bin
considered. We use a fourth-order interpolation and rewrite the cross section as
\begin{equation}
\sigma_{bin}=\sum_{p, k}\sum_{n_1, n_2, n_3}I_{k, p}
(n_1, n_2, n_3, \xi)(\alpha_s^p(\xi\mu_f)\Phi_{k})|_{n_1, n_2, n_3},
\end{equation}
where $\xi=\mu_r/\mu_f$, and $n_i$ are indices of grid points. The sum is performed over the
perturbative expansion order of QCD coupling constant ($p$) and the parton luminosities ($k$).
The parton luminosities involved here for up to NLO calculations are
\begin{align}   
\Phi_{1}&=\sum_{ij={qq'},{\bar q \bar q'}}x_1^2x_2^2f_{i/h_1}(x_1, \mu_f)f_{j/h_2}(x_2, \mu_f),\nonumber\\
\Phi_{2}&=\sum_{ij={qq},{\bar q \bar q}}x_1^2x_2^2f_{i/h_1}(x_1, \mu_f)f_{j/h_2}(x_2, \mu_f),\nonumber\\
\Phi_{3}&=\sum_{ij={q\bar q'}}x_1^2x_2^2f_{i/h_1}(x_1, \mu_f)f_{j/h_2}(x_2, \mu_f)+(h_1\leftrightarrow h_2),\nonumber\\
\Phi_{4}&=\sum_{ij={q\bar q}}x_1^2x_2^2f_{i/h_1}(x_1, \mu_f)f_{j/h_2}(x_2, \mu_f)+(h_1\leftrightarrow h_2),\nonumber\\
\Phi_{5}&=\sum_{ij={gq},{g\bar q}}x_1^2x_2^2f_{i/h_1}(x_1, \mu_f)f_{j/h_2}(x_2, \mu_f)+(h_1\leftrightarrow h_2).
\end{align}
Note that the $gg$ luminosity does not contribute to the cross sections induced by quark
contact interactions up to NLO. In the code the grid coefficients $I_{k, p}(n_1, n_2, n_3, \xi)$
with dependence on $\xi$ solved analytically, are calculated and stored in a table file for
each kinematic bin, which allows fast interpolations of $\sigma_{bin}$ for arbitrary PDF and
renormalization scale choices. The grid interpolation we used here is based on the original
approaches introduced by FastNLO~\cite{Kluge:2006xs} and APPLGRID~\cite{Carli:2010rw},
but with slightly modifications to adapt for our calculation.

\section{Program inputs}\label{s3}

The input parameters are specified in the files with the extension \gt{.card} in the
subdirectory \gt{data}. Each line contains a record for an input variable: a character
tag with the name of the variable, followed by the variable's value and descriptions.

\gt{proinput.card} contains various switches that

\begin{itemize}
\item
\gt{pdf} is the name of the PDF file read from LHAPDF~\cite{LHAPDF} used in the calculation, e.g.,
\gt{CT10.LHgrid} for CT10 NLO PDFs~\cite{Lai:2010vv}. Since the code has a fast PDF interpolation interface
in the final step, user can also change the PDF choice there. 
\item
\gt{pdfmember} specifies the PDF member used in the calculation as appearing in
LHAPDF, e.g., \gt{0} for central PDF.
\item
\gt{ang} specifies the two observables of the kinematic bin. Basically the program calculates
the total cross section in two-dimentional kinematic bins. For \gt{ang=1-3}, the first observable
of the bin is always the dijet invariant mass. The second observable
is $\chi=\exp(|y_1-y_2|)$ for \gt{ang=1}, $y^*=|y_1-y_2|/2$ for \gt{ang=2}, and
$y_{max}=\max(|y_1|, |y_2|)$ for \gt{ang=3}. While \gt{ang=4} is for single-inclusive jet
production with the first observable as $p_T$ of the individual jet and second as absolute rapidity
$|y|$ of the individual jet.
\item
\gt{mode=fitll/fitcs/fital} controls the calculation scheme. The code first calculates
the coefficients ($a, b$) defined in Eq.~(\ref{fit}) at both LO and NLO. The differences between
the three choices are that \gt{fitll} assumes only left-handed color-singlet operator at input scale
(coupling $c_1$). And \gt{fitcs} includes all the color-singlet operators at input scale
(couplings $c_1$, $c_3$, $c_5$). \gt{fital} considers all the six operators in Eq.~(\ref{opt}). User can choose
the calculation scheme according to the model studied since the most general case \gt{mode=fital}
consumes much more CPU times.   
\item
\gt{pseed} specifies the random-number seed used by the Monte Carlo integration.
\item
\gt{many} is the number of Monte Carlo sampling points specified for the calculation. User may need
to adjust it according to the numerical accuracy required. For an accuracy of a few per
mille, recommendation is \gt{3,000,000} or more.
\item
\gt{ppcollider} represents the type of the collider. \gt{0} for $p\bar p$ machine and \gt{1} for $pp$
machine.
\item
\gt{Sqrts} gives the center-of-mass energy $\sqrt s$ of the collider in GeV.
\item
\gt{scalescheme} specifies the choice of the hard momentum scale that defines the central
value of the factorization and renormalization scales. For dijet production (\gt{ang=1-3}),  
\gt{0} corresponds to choose the central value as average $p_T$ of the two leading jets, and
\gt{1} corresponds to choose the central value as $\exp(0.3y^*)$ times the harder jet $p_T$ in the
two leading jets. While for single-inclusive jet production, \gt{0} indicates
using the individual jet $p_T$ as the central value, and \gt{1} for using maximum
jet $p_T$ in the rapidity region as the central value. 

\end{itemize}

\gt{kininput.dat} contains parameters for clustering and selection of jets, in the same format
as in \gt{proinput.card}.
\begin{itemize}

\item
\gt{jetscheme} specifies the jet algorithm used in the calculations:
\gt{1} for the anti-$k_T$ jet algorithm~\cite{Cacciari:2008gp}, \gt{2} for the cone algorithms~\cite{Blazey:2000qt},
with \gt{Rsep=1.3} for modified Snowmass, and \gt{Rsep=2} for midpoint algorithm. Note here for the NLO
calculation the anti-$k_T$ algorithm is equivalent to the $k_T$~\cite{Catani:1993hr} or
Cambridge-Aachen~\cite{Dokshitzer:1997in} algorithm.
\item
\gt{recscheme} sets the recombination scheme~\cite{Salam:2009jx} used in jet clustering:
\gt{2} is for the 4-momentum scheme (recommended for LHC), and \gt{1} for the $E_T$ scheme.
\item
\gt{Rcone} sets the distance parameter or cone size for the jet algorithm used.
\item
\gt{Rsep} is only active for cone algorithm as explained above.
\item
\gt{ycut} and \gt{ptcut} specify conditions for jet acceptance, i.e., the maximal rapidity
and lowest $p_T$ (in GeV) for a cluster to be considered as a jet.
\item
\gt{yb, ys} specify the upper limit on $y_{boost}=|y_1+y_2|/2$ and $y^*$ of the dijet system,
respectively, and are not used for calculations of single-inclusive jet production. 
\end{itemize}

\gt{bininput.card} sets up a scan of the two-dimensional bin. Under tagged line \gt{massbin},
user can specify one invariant mass (or $p_T$) bin per line, containing the lower and upper limits (in GeV).
And under tagged line \gt{rapbin}, user can write one angular bin per line (corresponding
to $\chi$, $y^*$, $y_{max}$, or $|y|$ depending on \gt{ang}). The code will do a scan over
all the two-dimensional bins specified.

\section{Compilation and running}\label{s4}

After unzip the source file, go to the main directory and run \gt{make}. The executables will be
generated and moved into directory \gt{data}. The program uses the VEGAS subroutine from CUBA
library~\cite{Hahn:2004fe} for the Monte Carlo integration.

There are two main executables. \gt{dijets4ci\_sin} is for the calculation using a single CPU core.
And \gt{dijets4ci\_mul} is for the calculation using multicore which is much faster. Before 
runing \gt{dijets4ci\_mul}, user needs to first set the number of cores used by running
\gt{source setcores.sh \$1}, where \gt{\$1} is the number. Otherwise \gt{dijets4ci\_mul}
will use all the cores available. To run the code simply by entering
\gt{./dijets4ci\_sin \$tag} or \gt{./dijets4ci\_mul \$tag} under the
\gt{data} directory, where \gt{\$tag} specifies the name of the job. Note that during the
run the code needs to write into some auxiliary files for the parellel calculation. Also users
are suggested to use reasonable amount of cores since it costs
additional times for the master process to combine results from each cores.   
Besides, there are another two executables in \gt{fastCI} directory, i.e., \gt{ciconv} and
\gt{cixsec}, which will be explained later.

\section{Program outputs}\label{s5}  

At the end of each run, the program generates an output file \gt{\$tag\_fitresults.dat} containing the
calculated coefficients in Eq.~(\ref{fit}) for the specified PDF, 6 groups in total for each bin
(LO and NLO with 3 scale choices). Note that the value of $\mu_0$ for each bin are also written
into the output file.

At the same time, the gird files (one for each bin) are generated and stored in the
\gt{fastCI/fgrid} directory. Under \gt{fastCI} directory, user can first run
\gt{./ciconv \$1 \$2 \$3 \$4} to calculate the coefficients for arbitrary PDFs and
scale choices, where argument \gt{\$1} is the PDF file name, \gt{\$2} is the number of PDF member,
\gt{\$3} is the grid file name as in \gt{fgrid} directory, and \gt{\$4} is the specified
file name for the output. The calculated coefficients, 18 groups for each bin
(LO and NLO with 9 scale choices), are written into file \gt{\$4}. Using the
coefficients, as well as $\mu_0$ value in \gt{\$4}, user can calculate the cross sections
for arbitrary scale $\Lambda$ and Wilson coefficients $c_i$ according
to Eq.~(\ref{fit}). Or user can run \gt{./cixsec \$1 \$2} to get the cross sections
for $\Lambda$ and $\lambda_i$ values specified in file \gt{input}, where \gt{\$1} is the output
from \gt{ciconv}, and the cross sections are written into \gt{\$2}.  

\section{Sample results}\label{s6}

\begin{figure}[h]\centering
\includegraphics[width=0.38\textwidth]{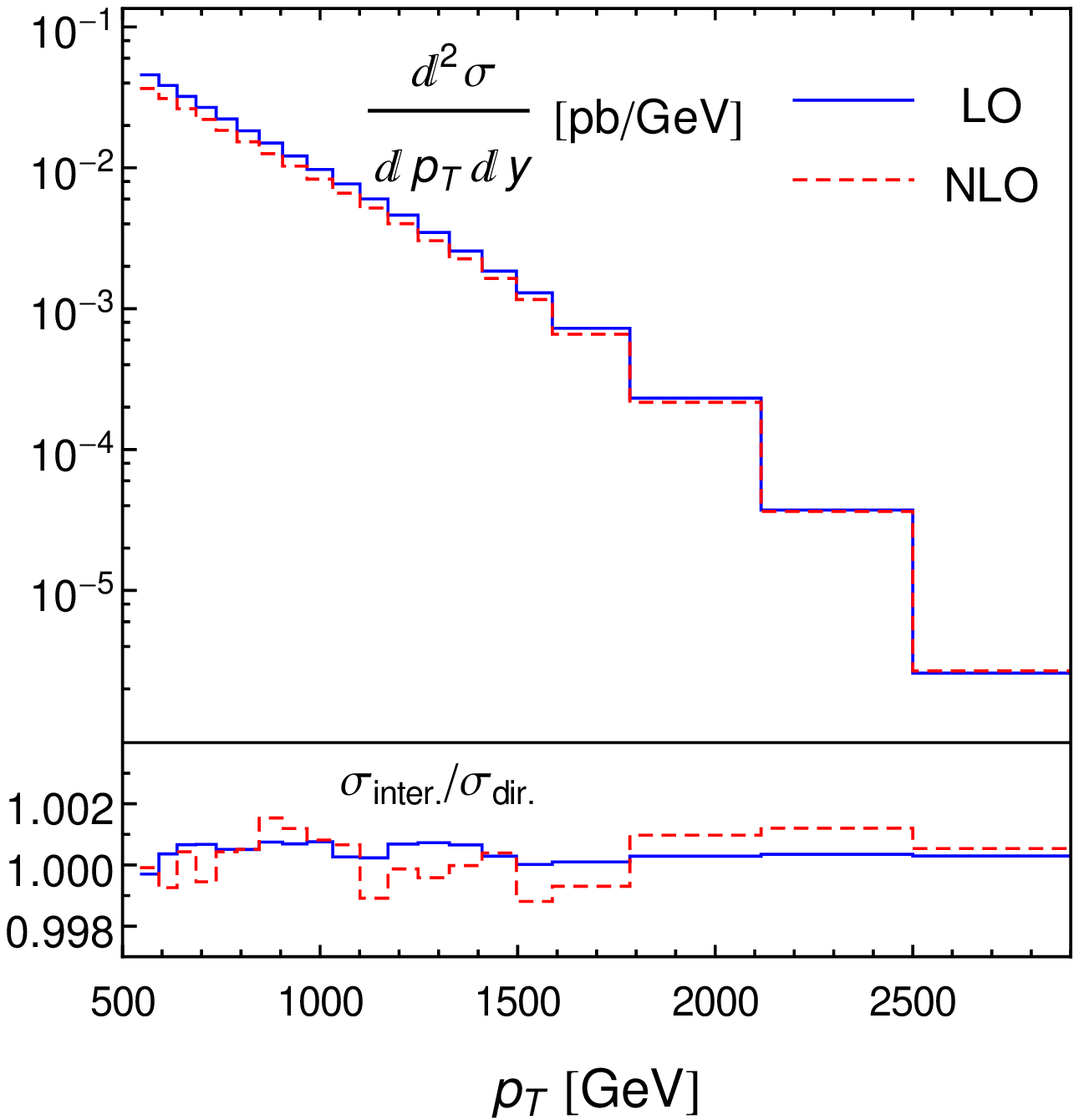}
\hspace{0.15in}
\includegraphics[width=0.38\textwidth]{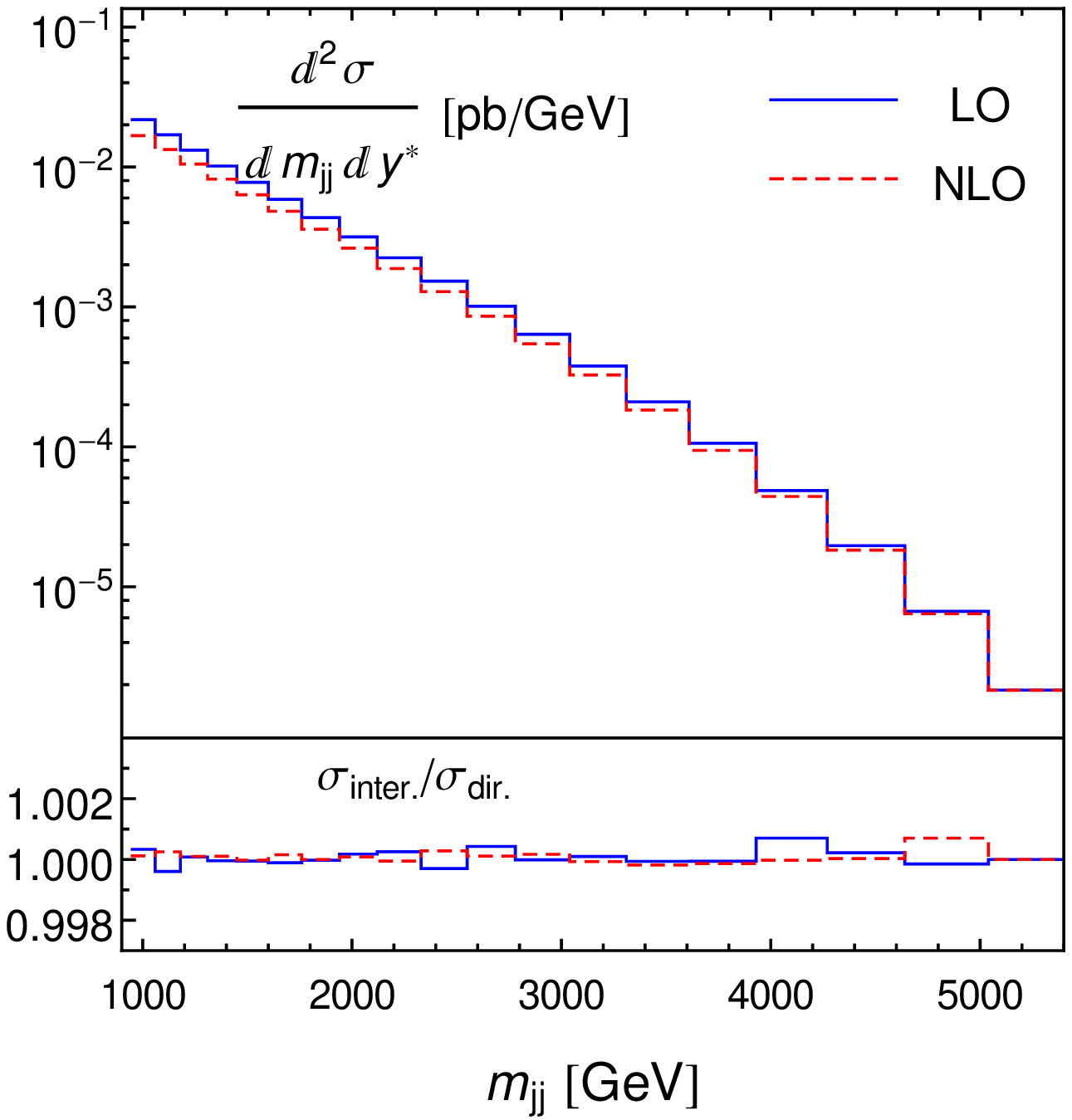}
\caption[]{Jet cross sections induced by the CIs at both LO and NLO, using the
central scale choice and CT10 NLO PDF. Ratios between the results directly from
the numerical integration and the results from latter interpolation using the weight grids
are also shown. Left and right panel correspond to single-inclusive jet and dijet
production respectively.} \label{ben}
\end{figure}

In this section we present some representative sample results from the program for both
single-inclusive jet and dijet production at LHC 8 TeV. For simplicity we just show
results for a single model with $\Lambda=7$ TeV and vector-like color-singlet coupling,
$\lambda_{1,5}=-1$, $\lambda_3=-2$, $\lambda_{2,4,6}=0$. We use the anti-$k_T$
jet algorithm with $R=0.7$ for single-inclusive jet production and $R=0.6$ for dijet
production. The \gt{scalescheme} is set to \gt{0} for single-inclusive jet production
and \gt{1} for dijet production. And we only present the differential cross sections
in the central rapidity region, i.e., $|y|<0.5$ or $|y^*|<0.5$ for single-inclusive jet
and dijet production, which is preferred by the contributions from CIs. In Fig.~\ref{ben}
we show comparisons of the LO and NLO jet cross sections. The NLO QCD corrections are found
to reduce the cross sections in most of the cases except for the extremely large $p_T$
or large invariant mass region. Also in Fig.~\ref{ben} we plot the ratios of the cross sections
that directly from the numerical integration and from latter interpolations. We can see very
good agreements between them, and the interpolation only introduces numerical errors of
the order 0.1\%, which are negligible in current analyses.

\begin{figure}[h]\centering
\includegraphics[width=0.38\textwidth]{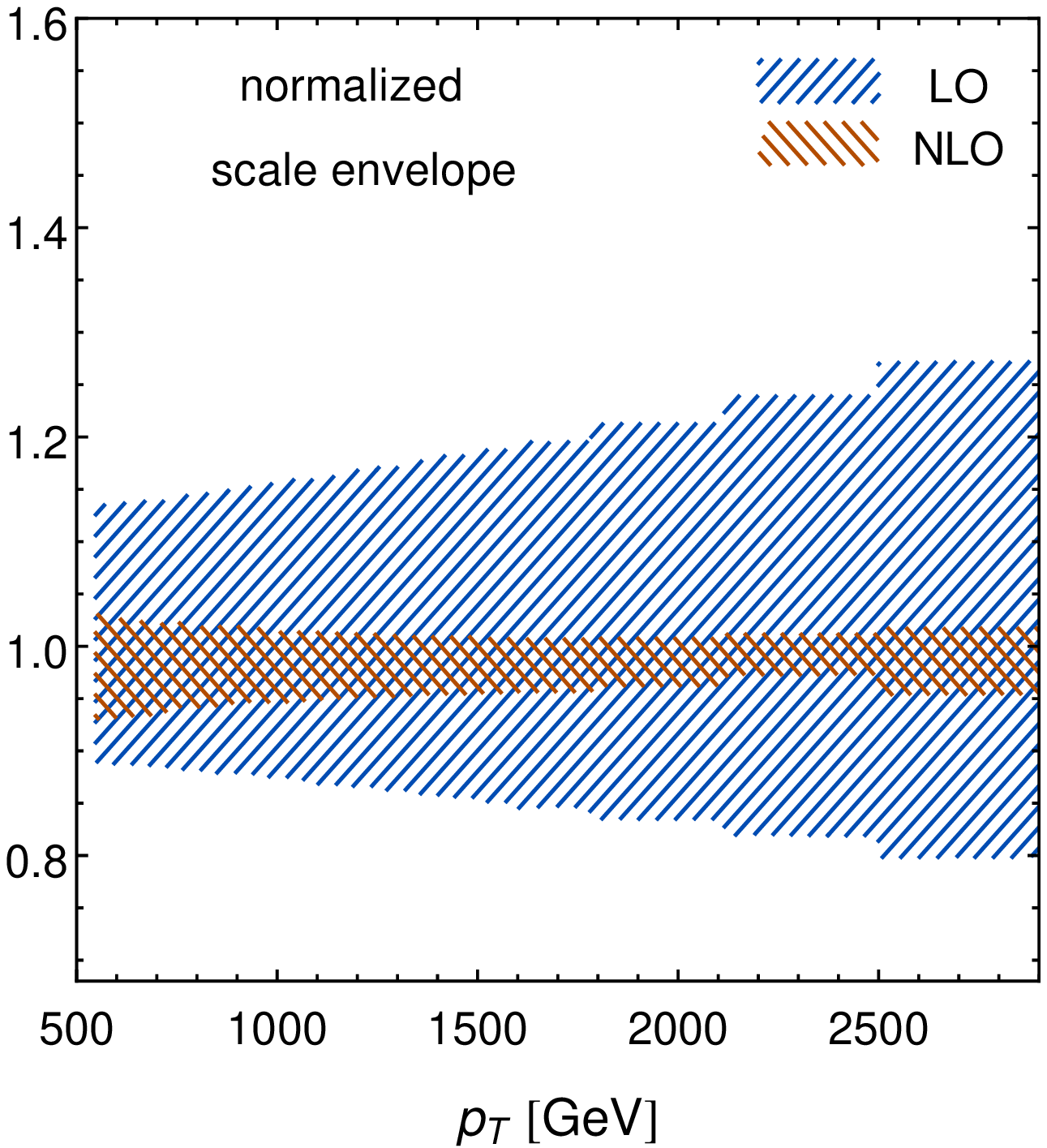}
\hspace{0.2in}
\includegraphics[width=0.38\textwidth]{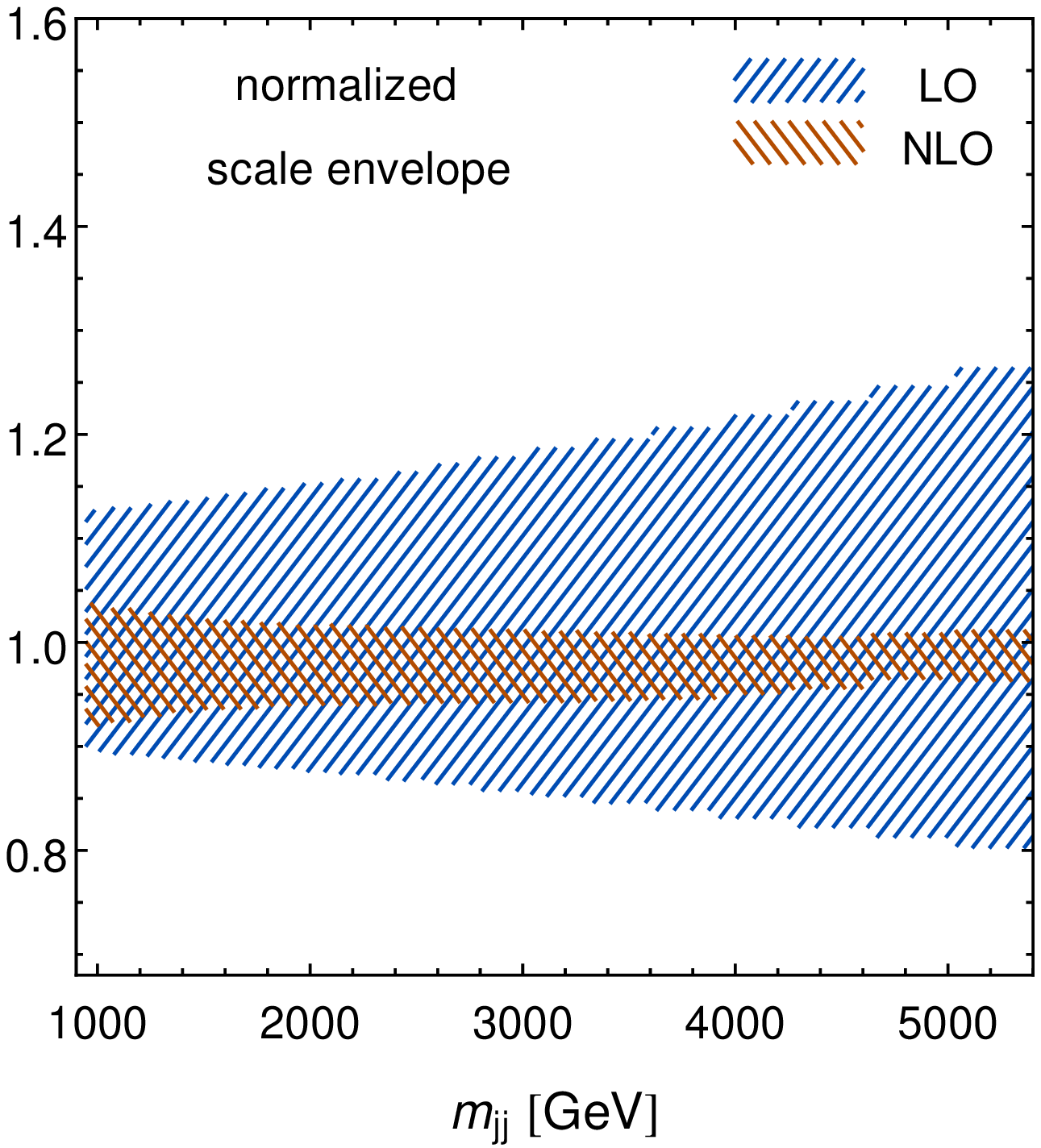}
\caption[]{Scale uncertainty envelope of the jet cross sections induced by the
CIs at both LO and NLO normalized to the central scale predictions, using the
CT10 NLO PDF. Left and right panel correspond to single-inclusive jet and dijet
production respectively.} \label{scale}
\end{figure}

\begin{figure}[h]\centering
\includegraphics[width=0.38\textwidth]{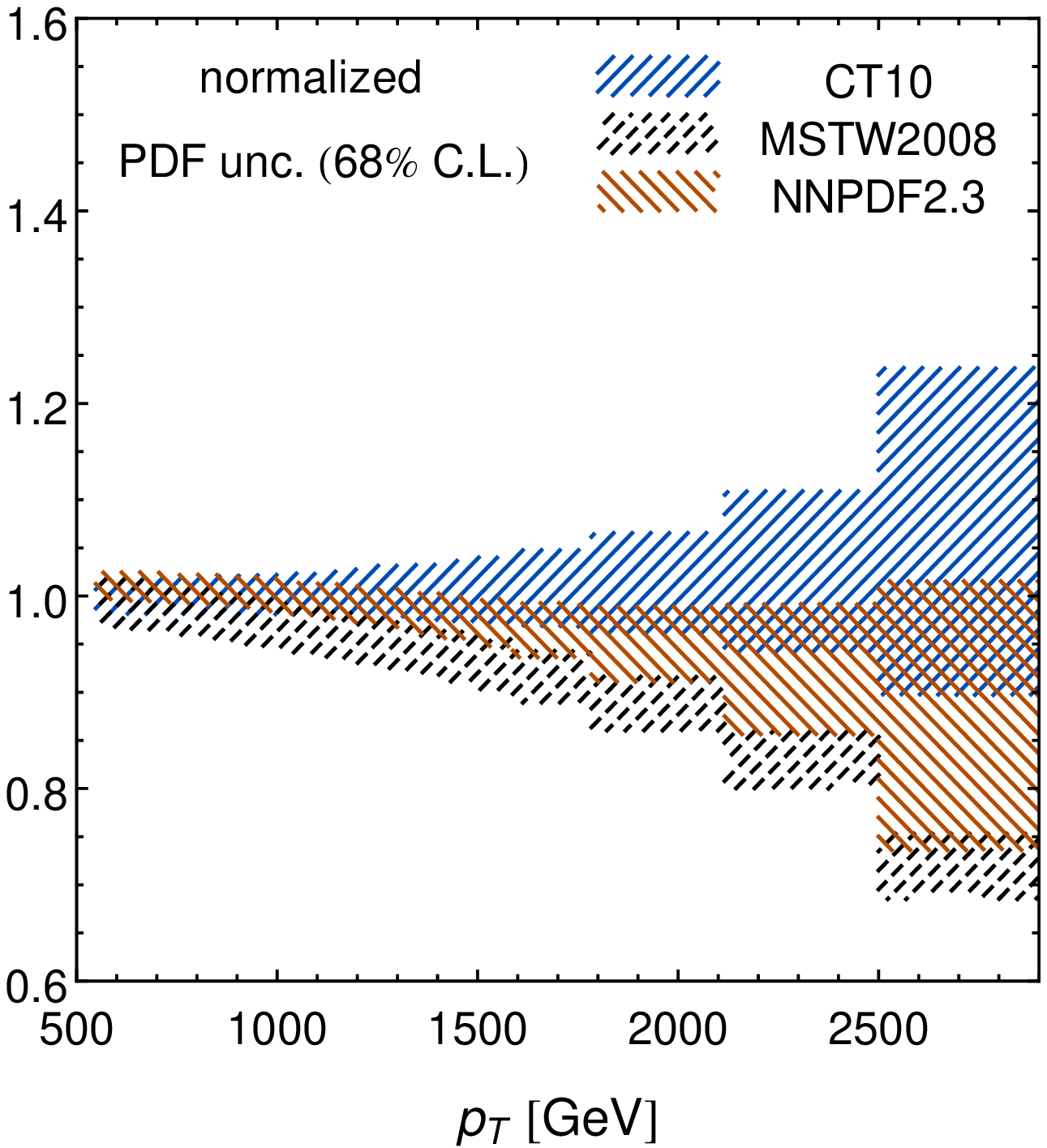}
\hspace{0.2in}
\includegraphics[width=0.38\textwidth]{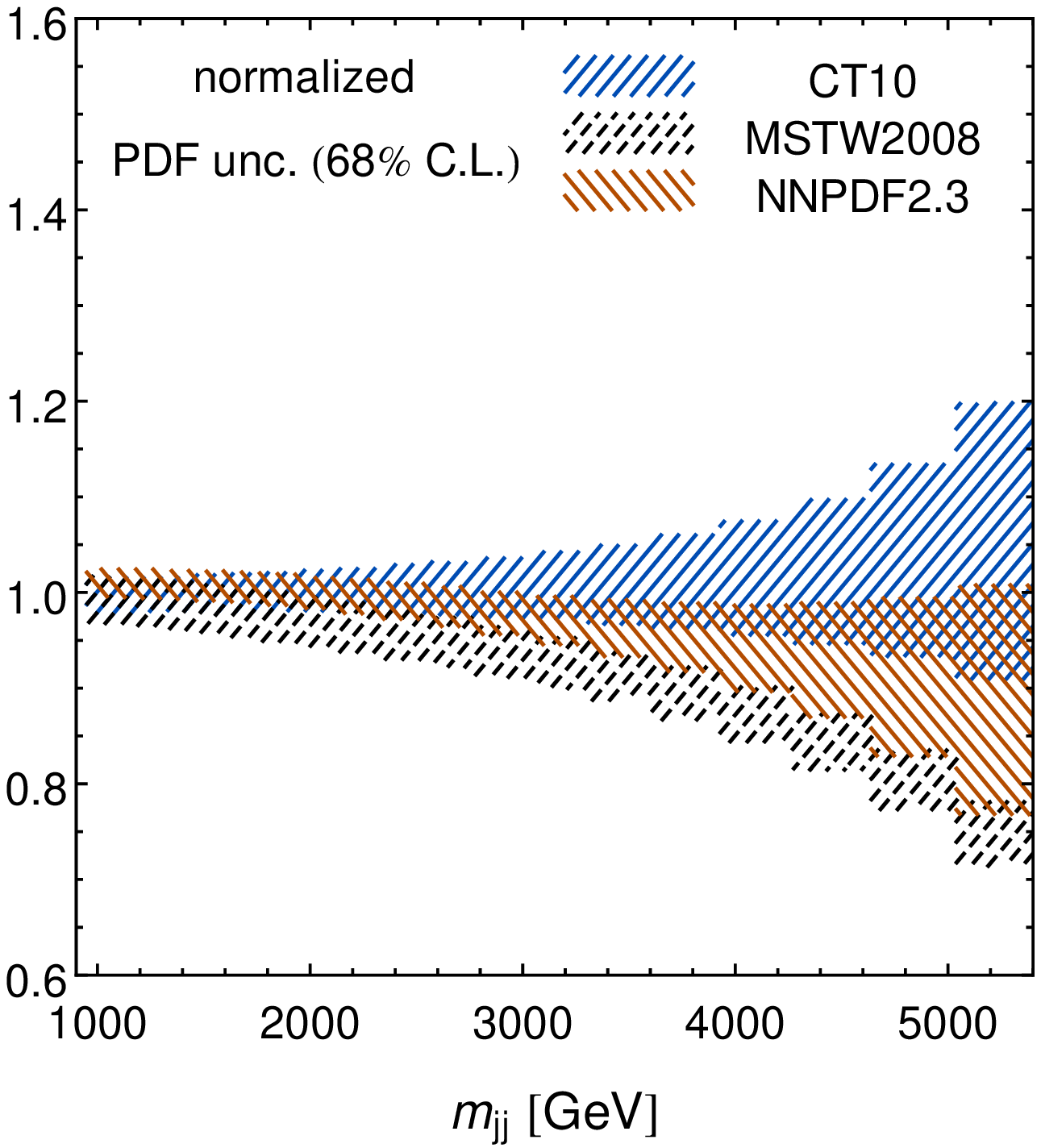}
\caption[]{PDF uncertainties of the jet cross sections induced by the CIs at
NLO normalized to the CT10 central predictions, using the central scale choice.
Results are shown for CT10, MSTW2008 and NNPDF2.3 NLO PDFs. Left and right
panel correspond to single-inclusive jet and dijet production respectively.}
\label{pdf}
\end{figure}

Using the fast interpolation we can obtain the results for arbitrary scale and PDF choices
without redoing the numerical integration, which is the most time consuming part.
In Fig.~\ref{scale} we plot the scale variation envelope of the jet cross sections by varying
the factorization and renormalization scales independently around the central value by a factor
of two. The cross sections are normalized to the LO or NLO predictions using the central
scale value. We can see that the NLO corrections greatly reduce the scale uncertainties,
especially the factorization scale uncertainties, for both cases, which makes the theoretical
predictions more reliable. Furthermore, in Fig.~\ref{pdf} we compare the NLO predictions with
68\% C.L. PDF errors from CT10~\cite{Lai:2010vv}, MSTW2008~\cite{Martin:2009iq}, and 
NNPDF2.3~\cite{Ball:2012cx} NLO PDFs. CT10 gives higher
predictions in both cases, and MSTW2008 shows much smaller PDF uncertainties in the extremely
large $p_T$ and large invariant mass region.

\section{Conclusions}\label{s7}
In conclusion, this document describes the CIJET1.0 program that aiming for the
calculation of single-inclusive jet or dijet production cross sections induced by quark
contact interactions from new physics at hadron colliders, up to NLO in QCD. It covers
various contact interactions with different chiral and color structures. The program
allows for parellel calculations of the numerical integrations using the CUBA library.
Moreover, the program includes a fast interpolation modular that could be used for
calculations with arbitrary parton distribution functions without redoing the time
consuming numerical integrations. The program could be used for the experimental
analyses of the quark compositeness at the LHC.

\section*{ACKNOWLEDGMENTS}
This work was supported by the U.S. DOE Early Career Research Award
DE-SC0003870 by Lightner-Sams Foundation. J. Gao appreciates helpful
discussions with Pavel Nadolsky, Harrison Prosper, Pavel Starovoitov, and
Klaus Rabbertz for the implementation of the grid interpolation.

\end{document}